\documentclass[twocolumn,showpacs,preprintnumbers,amsmath,amssymb,groupedaddress,prl]{revtex4}

\usepackage{graphicx}
\usepackage{dcolumn}
\usepackage{bm}
\usepackage{epsfig}

\def\e{\begin{equation}}
\def\f{\end{equation}}
\def\=#1{\overline{\overline #1}}

\def\-#1{{\bf #1}}
\def\.{\cdot}

\begin{document}

\title{Experimental study of the sub-wavelength imaging by a wire medium slab}

\author{Pavel A. Belov}
\affiliation{Department of Electronic Engineering, Queen Mary
University of London, Mile End Road, London, E1 4NS, United Kingdom}

\author{Yan Zhao}
\affiliation{Department of Electronic Engineering, Queen Mary
University of London, Mile End Road, London, E1 4NS, United Kingdom}

\author{Sunil Sudhakaran}
\affiliation{Department of Electronic Engineering, Queen Mary
University of London, Mile End Road, London, E1 4NS, United Kingdom}

\author{Akram  Alomainy}
\affiliation{Department of Electronic Engineering, Queen Mary
University of London, Mile End Road, London, E1 4NS, United Kingdom}

\author{Yang Hao}
\affiliation{Department of Electronic Engineering, Queen Mary
University of London, Mile End Road, London, E1 4NS, United Kingdom}

\begin{abstract}
An experimental investigation of sub-wavelength imaging by a wire
medium slab is performed. A complex-shaped near field source is used
in order to test imaging performance of the device. It is
demonstrated that the ultimate bandwidth of operation of the
constructed imaging device is 4.5\% that coincides with theoretical
predictions [Phys. Rev. E 73, 056607 (2006)]. Within this band the
wire medium slab is capable of transmitting images with $\lambda/15$
resolution irrespectively of the shape and complexity of the source.
Actual bandwidth of operation for particular near-field sources can
be larger than the ultimate value but it strongly depends on the
configuration of the source.
\end{abstract}

\pacs{78.20.Ci, 42.70.Qs, 42.25.Fx, 73.20.Mf}

\maketitle

\begin{figure}[h]
\centering \epsfig{file=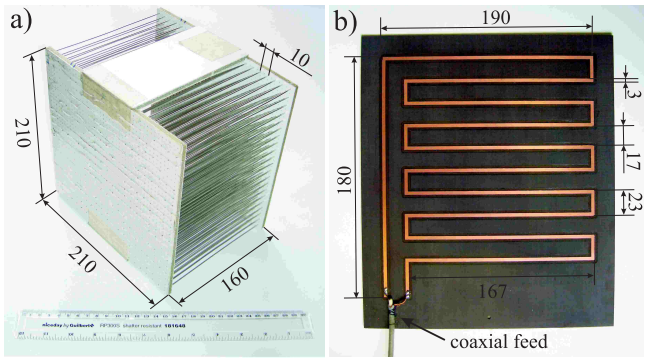, width=8.5cm} \caption{(Color
online) The geometries of the transmission device (a), an $21\times
21$ array of wires with $1$ mm radii, and the near-field source (b).
All dimensions are in millimeters.} \label{geom}
\end{figure}

An original concept of imaging with resolution smaller than the
wavelength (sub-wavelength imaging) has been recently proposed in
\cite{SWIWM}. It has been demonstrated that a regular array of
parallel conducting wires (see Fig. \ref{geom}(a)) is capable of
transporting images with sub-wavelength details from one planar
interface to another. The principle of operation of this device (see
\cite{canal}) is based on the idea of transforming the whole spatial
spectrum of the sub-wavelength source to the propagating modes
inside of a metamaterial formed by the array of wires, also called
as the wire medium \cite{WMPRB}. In such a way, the evanescent waves
which carry sub-wavelength information and normally decay in free
space can be transformed into the propagating modes inside of the
wire medium and transported to significant distances.

The initial experimental investigation of the sub-wavelength imaging
capability of the wire medium slab has been performed recently in
\cite{SWIWM}. The antenna in the form of letter `P' was used as a
sub-wavelength source. The clear images of the source were detected
at the back interface of the transmission device and resolution of
$\lambda/15$ was demonstrated for $18\%$ operation bandwidth. The
extensive theoretical studies \cite{resolWM} based on analysis of
transmission and reflection coefficient predicts that the
sub-wavelength imaging should be observed for at least $4.5\%$
bandwidth for any kind of the source. However, in practice, for
certain sources the imaging can be observed within larger frequency
bandwidths. The complexity of the near field produced by the source
and the interaction between the source and the transmission device
play crucial roles in determining the imaging performance of the
whole system. At the frequencies outside of the theoretical minimum
band of operation the strong reflections from the wire medium slab
are expected in accordance to \cite{resolWM}. That is why the
sensitivity of the source with respect to external fields becomes an
issue. If the source is very complex and contains a lot of
sub-wavelength details then its near field distribution can be
easily deformed by harmful reflections from the interface of the
transmission device and no proper sub-wavelength imaging can be
observed at the frequencies outside of the theoretical minimum band
of operation. However, if the source is simple and does not contain
many sub-wavelength details then the source is immune from
reflections  and the image can be successfully transported to the
back interface even at some frequencies outside of the minimum band
of operation, as it was observed in \cite{SWIWM}, for example.

\begin{figure*}[t]
\centering \epsfig{file=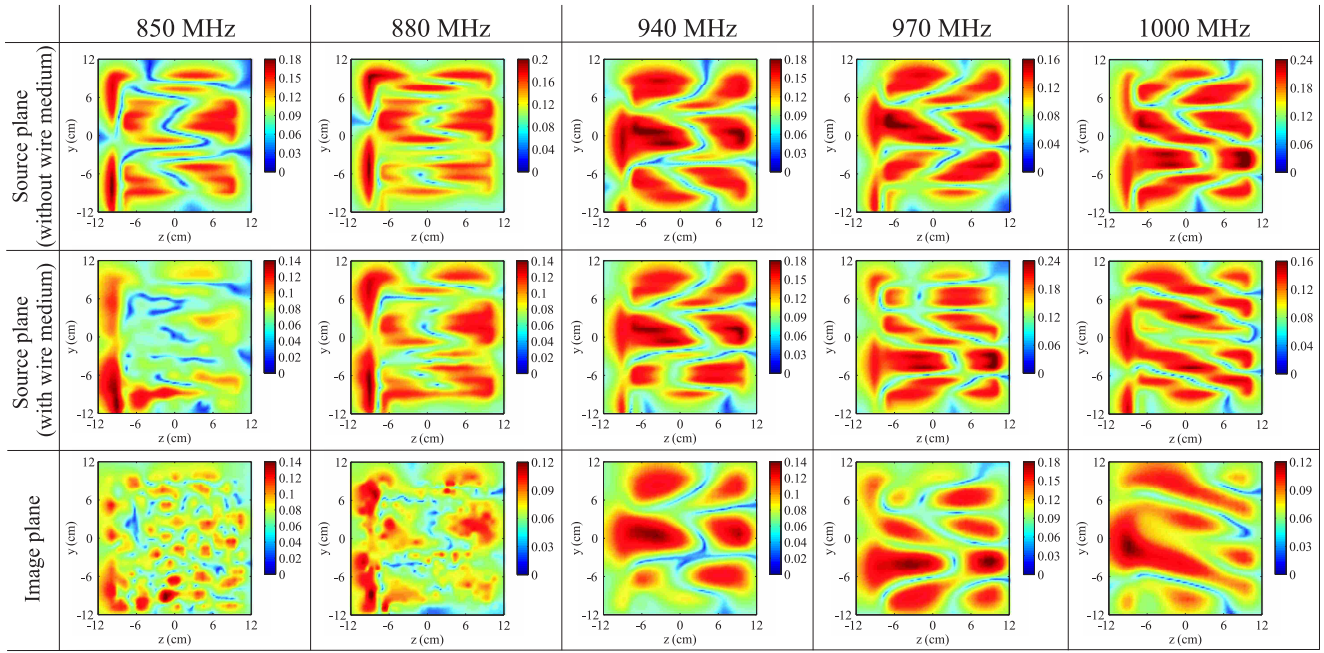, width=17.5cm}  \caption{(Color
online) Results of the near field scan at 850 MHz, 880 MHz, 940 MHz
and 1 GHz (in arbitrary units): the amplitude of the component of
electric field normal to the interface at 2 mm distance from the
meander antenna in the free space (source plane without wire
medium), the same but when the antenna is placed at the front
interface of the transmission device (source plane with wire medium)
and at 2 mm distance from the back interface of the wire medium slab
(image plane).\vspace{-3mm}} \label{meas}
\end{figure*}

\begin{figure}[b]
\vspace{-3mm} \centering \epsfig{file=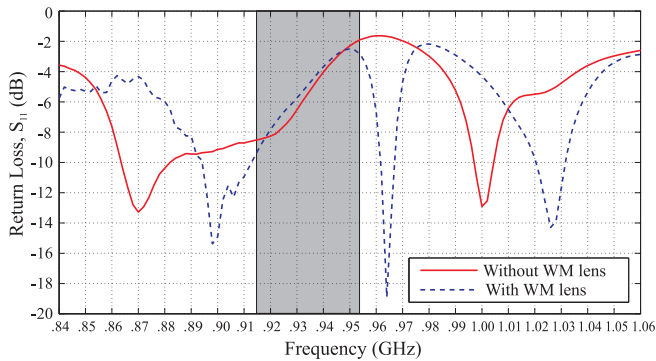, width=8.6cm}
\caption{(Color online) The return loss ($S_{11}$ parameter) as
function of frequency for the meander-line antenna in the free space
and at the interface of the wire medium slab.}\label{s11}
\end{figure}

In order to investigate the imaging capability of the wire medium
slab in details we have performed an experiment with the
meander-line antenna printed on 2 mm thick slab of duroid with
relative permittivity $\varepsilon=2.33$ (see Fig. \ref{geom}(b) for
other dimensions), which intentionally has much more complex
near-field distribution as compared to the `P' antenna used in
\cite{SWIWM}, see Fig. \ref{meas}. The return loss ($S_{11}$
parameter) within the frequency band from 840 MHz to 1060MHz for the
meander-line antenna in the free space was compared with the return
loss of the same antenna but placed close to the front interface of
the wire medium slab, see Fig. \ref{geom}(a). The results of the
comparison are presented in Fig. \ref{s11}
and clearly demonstrate that the wire medium slab does not affect
the meander-line antenna at the frequency band from 915 MHz to 955
MHz, see the shaded area in Fig. \ref{s11}. It means that within
915-955 MHz frequency range the meander-line antenna practically
does not suffer from reflections from the wire medium slab. The slab
is practically transparent at these frequencies and this fact was
predicted theoretically in \cite{resolWM} where unprecedentedly
small values of reflection coefficient for all angles of incidence
including evanescent waves have been demonstrated.

The near field scan has been performed for the frequencies within
840-1060 MHz frequency band which is significantly wider than the
band of 915-955 MHz where perfect imaging is expected in order to
verify general behavior of the imaging system. We used an automatic
mechanical near-field scanning device and a 2 mm long monopole probe
made from the central core of a coaxial cable with 2 mm diameter.
The scan area was $24\times 24$ cm$^2$ with 75 steps in both
directions. The probe was oriented normally with respect to the
interfaces of both the meander antenna and the transmission device.
So, it detected only the normal component of electrical field. The
wire medium slab is capable of imaging only the electromagnetic
waves with TM (transverse magnetic) polarization \cite{SWIWM} and
only the normal component of electric field is completely restored
at the back interface. The other two components contain contribution
of electromagnetic waves with TE (transverse electric) polarization,
which are not transferred by the wire medium slab.

The slab of wire medium is a transmission device, not a usual lens.
It transports electric field from its front interface to the back
interface and does not involve any focusing effects. The electric
field at 2 mm distance from the front interface of the meander-line
antenna located in free space (without the wire medium) was scanned
and regarded as the source field. After that the meander-line
antenna was placed at the front interface of the wire medium slab
and the field at 2 mm distance from the front interface of the
antenna was scanned once gain. This allows us to detect the
difference between the field created by antenna with and without the
presence of wire medium. The image field was scanned at 2 mm
distance away from the back interface of the slab in order to avoid
touching of the probe and the transmission device. Results of the
near-field scan at 23 frequencies from 840 MHz to 1060 MHz with 10
MHz step are presented in the multimedia file \cite{movie}. The same
results, but only for 850 MHz, 880 MHz, 940 MHz and 1 GHz are shown
in Fig.~\ref{meas}. At 910-960 MHz the fields at the source plane
with and without presence of antenna are practically identical (see
\cite{movie} or Fig. \ref{meas} for result at 940 MHz). This
confirms that the wire medium slab practically does not introduce
reflections at theses frequencies. At the same time the field in the
image plane repeats the source field with accuracy about 2 cm. This
confirms that the resolution of the imaging device at this
frequencies is about $\lambda/15$.

At frequencies lower than 920 MHz (up to 870 MHz) the fields in the
source plane with and without wire medium slab remain practically
identical. However, the image is distorted by sharp maxima (see
\cite{movie} or Fig. \ref{meas} for result at 880 MHz). These maxima
are caused by surface waves excited at the interfaces of the wire
medium and were predicted theoretically in \cite{resolWM}. We can
say that at these frequencies the transmission device maintains the
capability of sub-wavelength imaging, but with reduced resolution.
At frequencies lower than 870 MHz the surface waves completely
degrade the image and simultaneously provide strong reflections
which make distribution in the source and image planes different.

At frequencies higher than 960 MHz the imaging performance of the
transmission device also degrades, but this happens because of other
reasons. At 970 MHz the distributions at the source plane with and
without wire medium already become significantly different. It can
be explained by strong reflections from the slab which change
distribution of currents in the antenna. In this case reflections
are much more prominent than those at lower frequencies and are
caused by the fact that the slab does not fulfill the Fabry-Perot
resonance condition anymore. Following the theoretical studies
\cite{resolWM}, at lower frequencies the reflection coefficient is
large only for spatial harmonics with wave vectors close to the wave
vector of the surface wave. That is why while the wave vector of the
surface wave is large (870-920 MHz) we observe only sharp maxima in
the image plane and no significant changes between fields in the
source plane with and without wire medium. As the wave vector of the
surface wave decreases ($ <870$ MHz) the reflections experienced by
the antenna from the wire medium increase and completely destroy the
imaging. In the case of high frequencies ($> 960$ MHz) there are no
surface waves, but the reflection increases and this increase
happens simultaneously for all spatial harmonics. That is why we
observe strong difference between fields in the source plane with
and without wire medium at these frequencies. However, it is
interesting to note that the distributions in the source and image
planes of the transmission device remain practically identical (see
\cite{movie} or Fig. \ref{meas} for result at 970 MHz and 1 GHz).
The difference is practically negligible at 960-1060 MHz. The
resolution remains the same (2 cm, about $\lambda/15$) as at lower
frequencies. Following the theoretical predictions \cite{resolWM}
the resolution should slightly degrade with an increase of
frequency, however, within the tested frequency range we were not
able to detect any significant degradation of resolution.

Thus, we can conclude that the wire medium slab has good
sub-wavelength imaging properties even at frequencies higher than
the frequency of Fabry-Perot resonance, but the large level of
reflections from the wire medium slab is an issue. If the
sub-wavelength source is sensitive to external field (for example,
the meander-line antenna whose current distribution changes in an
external field), then the wire medium slab can not be used for its
imaging. However, if the source field is not sensitive to external
fields (for example, an array of small antennas fed by fixed current
sources), then it remains unaffected by reflections from the
transmission device and the wire medium slab can be successfully
used for imaging of this source with very good sub-wavelength
resolution. The antenna in the form of P-letter used in \cite{SWIWM}
is insensitive to reflections from the wire medium slab and that is
why the sub-wavelength imaging with $\lambda/15$ resolution was
reported in \cite{SWIWM} for the range from 920 MHz to 1.1 GHz.

In conclusion, in this letter the minimum operation bandwidth of the
wire medium slab as the sub-wavelength imaging device, theoretically
predicted in \cite{resolWM}, was confirmed experimentally. We would
like to stress that the actual bandwidth of operation significantly
depends on the complexity and sensitivity of the source to
reflections from the wire medium slab. For sources which are not
sensitive to external fields the sub-wavelength imaging can be
performed within significantly wide frequency range. However, for an
arbitrary source the imaging can be guaranteed only within the
minimum bandwidth.  We would like to remind \cite{SWIWM,resolWM}
that the wire medium slabs are able to transmit images to any long
distances specified by particular application. The only restriction
is that the length of the transmission device should be equal to an
integer number of half-wavelengths in order to fulfill Fabry-Perot
condition and eliminate unwanted reflections. The resolution of the
wire medium slab is ultimately defined by its period. That is why in
the microwave frequency range practically any fine sub-wavelength
resolution can be obtained if the wire medium with the sufficiently
small period can be manufactured.
\newpage
\end{document}